\documentclass[aps,amsmath,nofootinbib,superscriptaddress,amssymb,preprintnumbers,floatfix,showpacs,preprint]{revtex4-1}
\usepackage{amsmath,txfonts,longtable,booktabs,overpic,amssymb,bm,bbm,multirow,float,graphicx,color,dcolumn,subfigure,hyperref,slashed,feynmp}
\usepackage{tikz}
\usetikzlibrary{arrows,decorations.pathmorphing,decorations.markings,backgrounds,positioning,fit,petri,mindmap,spy}
\definecolor{blue}{RGB}{45,48,146}
\hypersetup{colorlinks,citecolor=black,anchorcolor=black,menucolor=black, linkcolor=black,filecolor=black,runcolor=black,urlcolor=black,frenchlinks=true}

\begin{document}

\title{Asymmetric pulse effects on pair production in chirped electric fields}
\author{Neng-Zhi Chen}
\affiliation{Key Laboratory of Beam Technology of the Ministry of Education, and College of Nuclear Science and Technology, Beijing Normal University, Beijing 100875, China}
\author{Orkash Amat}
\affiliation{Key Laboratory of Beam Technology of the Ministry of Education, and College of Nuclear Science and Technology, Beijing Normal University, Beijing 100875, China}
\author{Li-Na Hu}
\affiliation{Key Laboratory of Beam Technology of the Ministry of Education, and College of Nuclear Science and Technology, Beijing Normal University, Beijing 100875, China}
\author{Hong-Hao Fan}
\affiliation{Key Laboratory of Beam Technology of the Ministry of Education, and College of Nuclear Science and Technology, Beijing Normal University, Beijing 100875, China}
\author{Bai-Song Xie \footnote{bsxie@bnu.edu.cn}}
\affiliation{Key Laboratory of Beam Technology of the Ministry of Education, and College of Nuclear Science and Technology, Beijing Normal University, Beijing 100875, China}
\affiliation{Institute of Radiation Technology, Beijing Academy of Science and Technology, Beijing 100875, China}
\date{\today}
\begin{abstract}
We investigate the effects of the asymmetric pulse shapes on electron-positron pair production in three distinct fields: chirp-free, small frequency chirp, and large frequency chirp fields via the real-time Dirac-Heisenberg-Wigner formalism. Our findings reveal the disappearance of interference effects with shorter falling pulse length, and the peak is concentrated on the left side of the momentum spectrum. As the falling pulse length extends, an incomplete multi-ring structure appears in the momentum spectrum. The number density of particles are very sensitive to the asymmetry of the pulse. With a long falling pulse, the number density can be significantly enhanced by over four orders of magnitude when certain frequency chirps are utilized. These results highlight the impact of the effective dynamically assisted mechanism and the frequency chirp on pair creation.
\end{abstract}
\pacs{12.20.Ds, 11.15.Tk}
\maketitle

\section{Introduction}

Electron-positron ${(e^{-}e^{+})}$ pair production in the vacuum under strong electromagnetic fields stands as one of the most intriguing phenomena in relativistic quantum physics \cite{dirac1928quantum,sauter1931verhalten,anderson1933positive,heisenberg1936folgerungen,schwinger1951gauge,xie2017electron}. This effect has yet to be directly observed in the laboratory. The reason is that the critical field strength $E_{\mathrm{cr}}=m^2c^{3} / e \hbar \approx 1.3 \times 10^{16}~\rm {\mathrm{V}/\mathrm{cm}}$ is too high, which corresponds to the laser intensity of about $4.3\times10^{29}~\mathrm{W/cm^{2}}$ (where $m$ and $-e$ is the electron mass and charge, respectively). Such an intensity level is presently unattainable. Nevertheless, X-ray Free Electron Laser systems approach near-critical field strengths of ${E/{E}_{\mathrm{cr}}\approx 0.01-0.1}$ \cite{ringwald2001pair}. Recent advancements in high-intensity laser technology \cite{heinzl2009exploring,pike2014photon} provide prospects for potential experimental validation  in the near future.

The influence of different pulse profiles on pair production has been a subject of extensive research. Hebenstreit \textit{et al.} noted a significant sensitivity of momentum spectra to external field parameters using short-pulse lasers \cite{PhysRevLett.102.150404}. Furthermore, in spatially inhomogeneous pulse fields, the pair production process is accompanied by particle self-focusing effects \cite{PhysRevLett.107.180403}. Schützhold \textit{et al.} introduced a dynamically assisted Schwinger mechanism, which combines low-frequency strong fields with high-frequency weak fields \cite{schutzhold2008dynamically}, substantially amplifying the particle production rate.
Recently, the impact of frequency chirp effects on particle momentum spectra and number density in time-dependent  electric field \cite{dumlu2010schwinger,olugh2019pair,bai2022enhancement,wang2023enhancement} has attracted increased attention. Specifically, research has focused on enhancing particle creation in both time-dependent monochromatic and bichromatic laser fields subjected to frequency chirp \cite{abdukerim2017enhanced,ababekri2020chirp,wang2021effect,gong2020electron,mohamedsedik2021schwinger,LiLieJuan2021}, and exploring asymmetric pulse shapes in single-color fields \cite{oluk2014electron,olugh2020asymmetric}.
Numerous studies aim to enhance pair production through various field combinations.

In this study, we explore vacuum pair production, considering a range of chirp parameters and asymmetric pulse shapes in the electric field. Our approach is based on the Dirac-Heisenberg-Wigner (DHW) formalism, which can handle diverse chirp and asymmetric envelope fields. We observe that the momentum spectrum exhibits a high sensitivity to chirp, revealing interference effects for varying pulse shapes and chirp parameters. The particle number density is in general enhanced with increasing chirp. Specifically, in instances of extended falling pulses, the effect of chirp results in an enhancement of the particle number density by up to four orders of magnitude. Throughout our study, we adopt natural units ($\hbar = c = 1$) and express all quantities in terms of the electron mass $m$.

The paper is structured as follows: In Sec. \ref{Dirac-Heisenberg-Wigner formalism}, we provide a brief overview of the DHW formalism applied in this study. In Sec. \ref{field}, we introduce the model for background fields. Within Sec. \ref{momentum}, we present numerical outcomes concerning the momentum spectrum and analyze the underlying physics. In Sec. \ref{density}, we focus on numerical findings related to the number density. Finally, in Sec. \ref{conclusion}, we provide a summary and discussion.

\section{DHW formalism}\label{Dirac-Heisenberg-Wigner formalism}

The DHW formalism is an approach used to describe quantum phenomena within a system, utilizing the Wigner function to represent the relativistic phase space distribution. We employ the widely adopted DHW formalism for investigating vacuum pair production within strong background fields \cite{bialynicki1991phase,hebenstreit2010schwinger,hebenstreit2011particle,kohlfurst2020effect,ababekri2020chirp,li2021study,mohamedsedik2023phase,PhysRevD.108.056011, PhysRevD.107.116010}. Since the specific derivation of DHW has been explained in prior studies \cite{kohlfurst2015electron,vasak1987quantum}, this work focuses on presenting the fundamental concepts and essential aspects of this method.

For convenience, we start with the system's gauge-invariant density operator,
\begin{equation}\label{DensityOperator}
 \hat {\mathcal C}_{\alpha \beta} \left(r , s \right) = \mathcal U \left(A,r,s
\right) \ \left[ \bar \psi_\beta \left( r - s/2 \right), \psi_\alpha \left( r +
s/2 \right) \right],
\end{equation}
here, the electron's spinor-valued Dirac field is denoted by ${\psi }_{\beta }\left ( {x}\right )$, where $r$ represents the center of mass and $s$ denotes the relative coordinates. The introduction of the Wilson line factor, $\mathcal{U}(A,r,s) $, aims to maintain the gauge invariance of the density operator. This factor's properties are determined by the elementary charge $e$ and the background gauge field $A$.

An essential component of the DHW formalism is the covariant Wigner operator, which is obtained by Fourier transforming the density operator Eq.\eqref{DensityOperator},
\begin{equation}\label{WignerOperator}
\hat{\mathcal W}_{\alpha \beta} \left( r , p \right) = \frac{1}{2} \int d^4 s \
\mathrm{e}^{\mathrm{i} ps} \  \hat{\mathcal C}_{\alpha \beta} \left( r , s
\right).
\end{equation}
Upon calculating the vacuum expectation value of the Wigner operator, we obtain the Wigner function,
\begin{equation}\label{Wigner function}
 \mathbb{W} \left( r,p \right) = \langle \Phi \vert \hat{\mathcal W} \left( r,p
\right) \vert \Phi \rangle.
\end{equation}
To derive the time-evolution equation, we rely on the equal-time Wigner function
\begin{equation}
\mathbbm{w} (\mathbf{x}, \mathbf{p}, t) = \int \frac{d p_{0}}{2 \pi} \mathbb{W}(r, p).
\end{equation}

The Wigner function can be decomposed into a complete basis set of Dirac matrices, resulting in 16 covariant real Wigner components
\begin{equation}\label{decomposed}
\mathbbm{w} = \frac{1}{4} \left( \mathbbm{1} \mathbbm{s} + \textrm{i} \gamma_5
\mathbbm{p} + \gamma^{\mu} \mathbbm{v}_{\mu} + \gamma^{\mu} \gamma_5
\mathbbm{a}_{\mu} + \sigma^{\mu \nu} \mathbbm{t}_{\mu \nu} \right),\
\end{equation}
where  $\mathbbm{s}$, $\mathbbm{p}$, $\mathbbm{v}_{\mu}$, $\mathbbm{a}_{\mu}$ and $\mathbbm{t}_{\mu \nu}$ denote scalar, pseudoscalar, vector, axial vector and tensor, respectively. According to the  Refs. \cite{hebenstreit2010schwinger,hebenstreit2011particle,kohlfurst2015electron}, the motion equation for the Wigner function is:
\begin{equation}
D_{t}\mathbbm{w} = -\frac{1}{2}\mathbf{D}_{\mathbf{x}}[\gamma^{0}\bm{\gamma},\mathbbm{w}]
+\mathrm{i}m[\gamma^{0},\mathbbm{w}]-\mathrm{i}\mathbf{P}\{\gamma^{0}\bm{\gamma},\mathbbm{w}\},
\label{motion}
\end{equation}
where $D_{t}$, $\mathbf{D}_{\mathbf{x}}$ and $\mathbf{P}$ are represented as pseudodifferential operators.

By embedding the motion Eq. \eqref{decomposed} into Eq. \eqref{motion}, we obtain a system of partial differential equations for the 16 Wigner components. Besides, for spatial homogeneous time-dependent electric fields  we can use the characteristic method \cite{blinne2016comparison} to replace the dynamical momentum ${\mathbf p}$ with the canonical momentum ${\mathbf q}$ via $ {\mathbf q} - e {\mathbf A} (t)$, Thus, the system of partial differential equations for the 16 Wigner components can be simplified to a set of 10 ordinary differential equations:
\begin{equation}
{\mathbbm w} = ( {\mathbbm s},{\mathbbm v}_i,{\mathbbm a}_i,{\mathbbm t}_i)
\, , \quad  {\mathbbm t}_i := {\mathbbm t}_{0i} - {\mathbbm t}_{i0}  \, .
\end{equation}
Due to the complexity of the motion equation of the Wigner function, we do not present a detailed derivation here, but it is available in the cited Refs.
\cite{hebenstreit2011particle,kohlfurst2015electron}. The relevant non-zero vacuum initial values are as follows:
\begin{equation}
{\mathbbm s}_{vac} = \frac{-2m}{\sqrt{{\mathbf p}^2+m^2}} \, ,
\quad  {\mathbbm v}_{i,vac} = \frac{-2{ p_i} }{\sqrt{{\mathbf p}^2+m^2}} \, .
\end{equation}
Hereafter, we denote the scalar Wigner function $f(q,t)$ using the single-particle momentum distribution function,
\begin{equation}
f({\mathbf q},t) = \frac 1 {2 \Omega(\mathbf{q},t)} (\varepsilon - \varepsilon_{vac} ),
\end{equation}
where $\Omega=\sqrt{m^2+{\mathbf p}^2}=\sqrt{m^{2}+(\mathbf{q}-e\mathbf{A}(t))^{2}}$ is the total energy of particle, $\varepsilon = m {\mathbbm s} + p_i {\mathbbm v}_i$ represents the phase space energy density, $m$ represents the electron mass.

To accurately compute the single-particle momentum distribution $f({\mathbf q},t)$, we refer to Ref. \cite{blinne2016comparison}. We introduce an auxiliary three-dimensional vector $\mathbf{v} (\mathbf{q},t)$  to facilitate the calculation:
\begin{equation}
\mathbf{v} (\mathbf{q},t) : = {\mathbbm v}_i (\mathbf{p}(t),t) -
(1-f({\mathbf q},t))  {\mathbbm v}_{i,vac} (\mathbf{p}(t),t) \, .
\end{equation}
Consequently, by solving the following ordinary differential equations, the single-particle momentum distribution function $f(q, t)$ can be obtained:
\begin{equation}
\begin{array}{l}
\dot{f}=\frac{e\mathbf{E}\cdot \mathbf{v}}{2\Omega},\\
\dot{\mathbf{v}}=\frac{2}{\Omega^{3}}[(e\mathbf{E}\cdot \mathbf{p})\mathbf{p}-e\mathbf{E}\Omega^{2}](f-1)-\frac{(e\mathbf{E}\cdot \mathbf{v})\mathbf{p}}{\Omega^{2}}-2\mathbf{p}\times \mathbbm{a}-2m\mathbbm{t},\\
\dot{\mathbbm{a}}=-2\mathbf{p}\times \mathbf{v},\\
\dot{\mathbbm{t}}=\frac{2}{m}[m^{2}\mathbf{v}+(\mathbf{p}\cdot \mathbf{v})\mathbf{p}],
\end{array}\label{eq3}
\end{equation}
where $f(\mathbf{q},-\infty)=0 \, , \, \mathbf{v}(\mathbf{q},-\infty)=\mathbbm{a}(\mathbf{q},-\infty)=\mathbbm{t}(\mathbf{q},-\infty)=0 $ are initial conditions, dot represents a total time derivative, $\mathbf{v}$ denotes current density, $\mathbbm{a}$ is spin density, $\mathbf{E}$ is  electric field, ${\mathbf p}$ represents kinetic momentum, $\mathbf{q}$ denotes canonical momentum, $e$ is the charge of particle, i.e., $|e|$ and $-|e|$ for positron and electron, respectively, $\mathbf{A}(t)$ is the vector potential of the external field.

Ultimately, the number density of created pairs, defined at the asymptotic limit as $t\rightarrow+\infty$, can be determined by integrating the distribution function $f(\mathbf{q},t)$:
\begin{equation}\label{14}
  n = \lim_{t\to +\infty}\int\frac{d^{3}q}{(2\pi)^ 3}f(\mathbf{q},t) \, .
\end{equation}

\section{The external field form}\label{field}

We study the  pair production in asymmetric frequency-chirped  electric fields. The external electric field we consider is presented as follows:
\begin{equation}\label{model2}
\mathbf{E}(t)=E_{0}\left[\exp\left(-\frac{t^2}{2{\tau_1}^2}\right)H(-t)+\exp\left(-\frac{t^2}{2{\tau_2}^2}\right)H(t)\right]\cos\left(bt^2+\omega t\right),
\end{equation}
where $ E_0$ represents the electric field amplitude, $\tau_{1}$ and $\tau_{2}$ denote the rising and falling pulse duration, H($t$) stands for the Heaviside step function, $\omega$ corresponds to the oscillation frequency of the electric field, $b$ is the chirp parameter. We introduce a time-dependent effective frequency: $\omega_{eff} = \omega+ bt$.

\begin{figure}[t]
\begin{center}
\includegraphics[width=\textwidth]{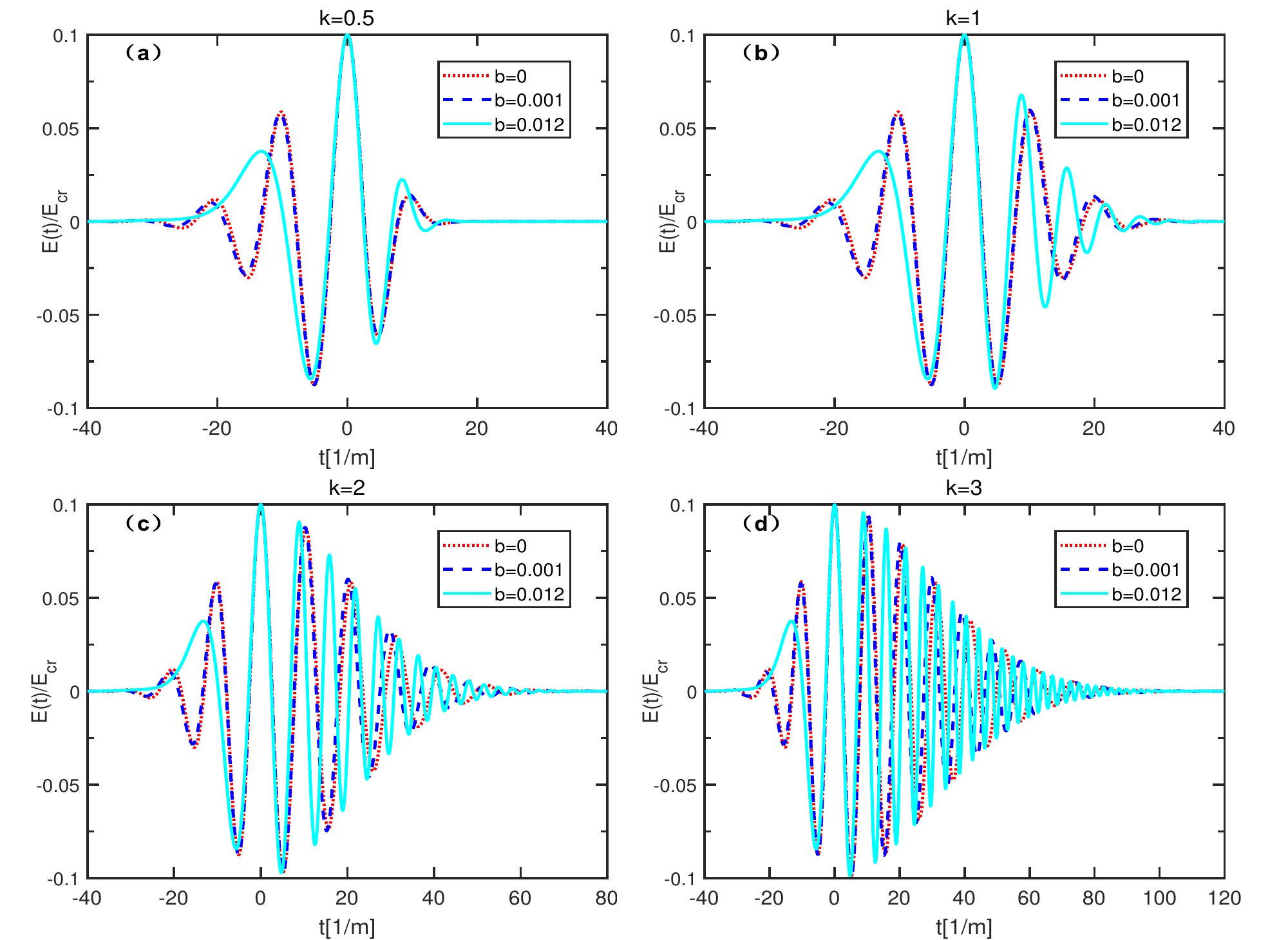}
\end{center}
\setlength{\abovecaptionskip}{-0.5cm}
\caption{Displaying asymmetric time-dependent electric field with ratio parameter $k = 0.5$, $k = 1$, $k = 2$ and $k = 3$. The red dashed line, purple dashed line and solid blue line are corresponding to the field with  chirp parameter $ b = 0 m^2$,  $ b = 0.001m^2$ and $ b = 0.012m^2$, respectively.
The chosen parameters are $E_{0}=0.1 E_{\mathrm{cr}}$, $\omega=0.6m$, $\tau_{1}=10/m$ and $\tau_{2}=k \tau_{1}$. }
\label{1}
\end{figure}

Notably, the electric field described by Eq.~\eqref{model2} is solely time-dependent and can be regarded as a standing wave formed by two laser beams with different pulse widths and opposite propagation directions. In our study of ${e^{-}e^{+}}$ pair production, we define the following electric field parameters:
\begin{equation}\label{eq:foobar}
E_{0}=0.1 E_{\mathrm{cr}}, \, \omega=0.6 \mathrm{~m}, \, \tau_{1}=10 / \mathrm{m},  \, \tau_{2}=k \tau_{1},
\end{equation}
where $ E_{\mathrm{cr}}=m^2c^3/e\hbar\approx1.3\times10^{16}\mathrm{V/cm}$ stands as the Schwinger critical field strength,
$k$ denotes the ratio of the falling pulse duration to the rising pulse duration, adjusting the field's asymmetry. In Fig. \ref{1}, we illustrate the influence of time-dependent electric field over time for different chirp pulse parameters $b$ and ratio parameter $k$.

In the study of vacuum pair creation, the Keldysh parameter holds significant importance and is defined as follows:
\begin{equation}\label{gamma}
\gamma = m\omega /eE,
\end{equation}
here, $E$ represents the field strength of the background field. The tunneling effect and multiphoton absorption can be investigated by considering $\gamma \ll 1$ and $\gamma \gg 1$, respectively. In our research, the Keldysh parameter is not significantly larger than 1, indicating that the creation of particles involves multiphoton absorption and tunneling effects. The pair production process primarily takes place within the time interval $ -\tau < t< \tau $, we select the chirp parameter to fall within the standard range, ensuring that $bt$ maximum value is approximately equal to $\omega$. The linear chirp parameter $b$ can be expressed as $b= \alpha \omega /\tau $, where $0\le \alpha \le 1$ with the maximum chirp value to be $ b = 0.012m^2$.

\section{Momentum spectra of the produced particles}\label{momentum}

This section explores the influence of asymmetric pulse shapes on  momentum spectra across varying fields, encompassing chirp-free, small frequency chirp, and large frequency chirp configurations.

\subsection{chirp-free $b = 0$}

\begin{figure}[t]
\begin{center}
\includegraphics[width=\textwidth]{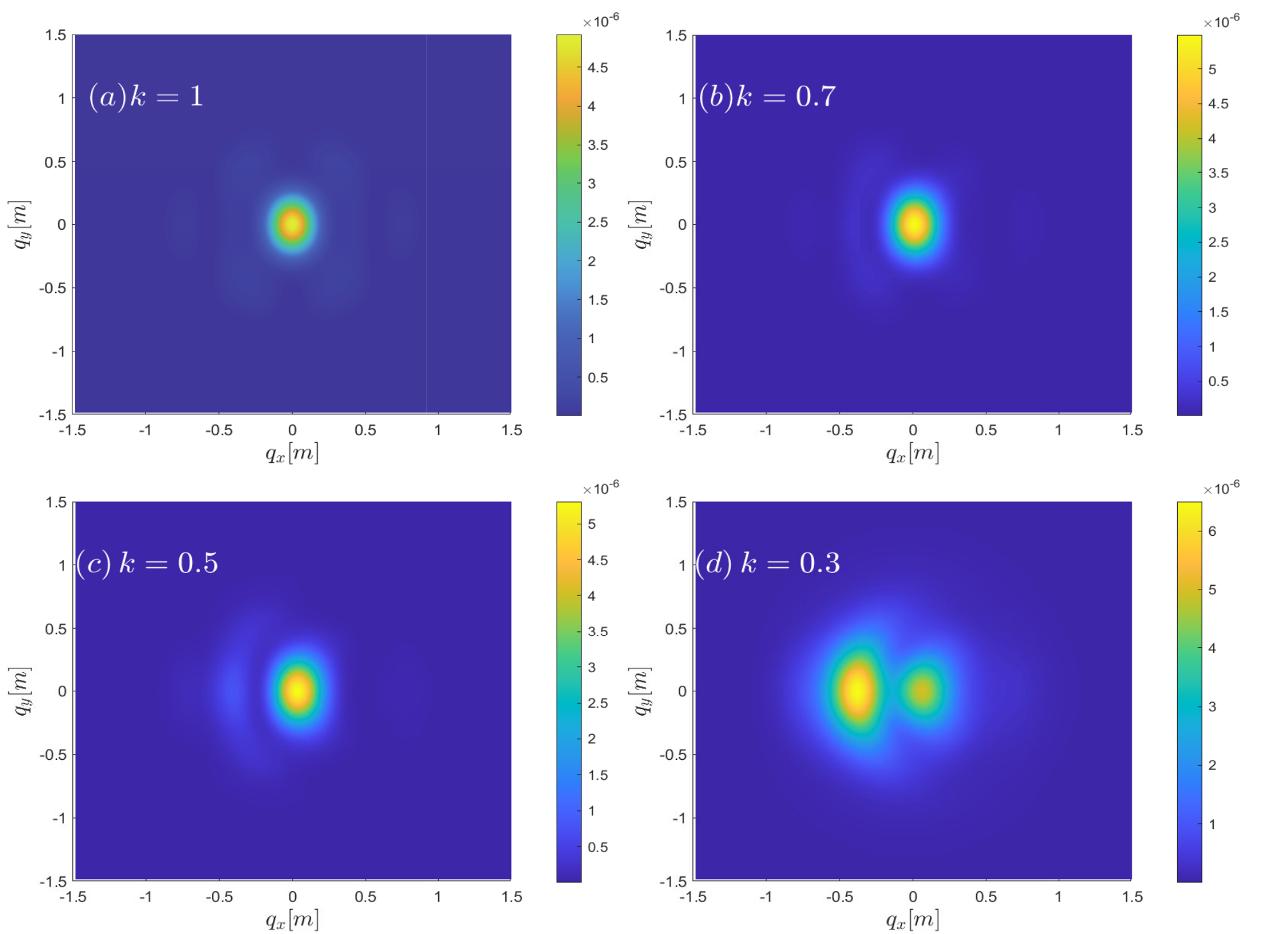}
\end{center}
\setlength{\abovecaptionskip}{-0.5cm}
\caption{Momentum spectra of particles in the $(q_x, q_y)$ plane with chirp-free $(b = 0)$. The falling-pulse duration $\tau_2 = k\tau_1$ is compressed, with $k$ varying between $0$ and $1$. The chosen parameters for the electric field are $E_0 = 0.1E_{\mathrm{cr}}$, $\omega = 0.6m$, and $\tau_1$ = $10/m$.}
\label{2}
\end{figure}

The impact of the asymmetric pulse on particle momentum spectra within the chirp-free field is illustrated in Fig. \ref{2}, exploring a range of $k$ values from $0$ to $1$. The momentum spectrum is  sensitive to the electric field asymmetry. At $k = 1$, the momentum spectrum is symmetric and peaked at the origin, exhibiting mild oscillations as depicted in Fig. \ref{2}(a). These oscillations arise from interference among separate complex conjugate pairs of turning points, detailed in Ref. \cite{dumlu2010schwinger}.
When  $ k = 0.7$, the peak has a slight increase with momentum concentrating around $-0.4m$ as shown in Fig. \ref{2}(b).  The breakdown of symmetry is a direct consequence of the asymmetric nature of the pulse. When setting $k$ to 0.5, the momentum spectra split in both positive and negative $q_x$ directions, resulting in the observation of two distinct maxima in the momentum spectrum in Fig. \ref{2}(c). This pronounced asymmetry in the momentum spectrum resembles the effects introduced by carrier phase, as studied in Ref. \cite{PhysRevLett.102.150404}. For $ k = 0.3$, the minor peak in the negative $q_x$ region grows into a major peak; see Fig. \ref{2}(d). This effect is similar to the frequency-chirp effect studied in Ref. \cite{olugh2019pair}. Finally, the maximum of the momentum spectrum rises from  $4.92\times 10^{-6} ( k = 1)$ to $ 6.82\times 10^{-6} ( k = 0.3)$.

For small values of $k$, the role of pulse asymmetry is similar to that of the carrier phase. In the presence of $E(t)$, the produced pairs are continuously accelerated, and particle momentum is mainly determined by their creation time \cite{orthaber2011momentum,kohlfurst2018phase}. Particles are created with zero longitudinal momentum at the earlier time $t_0$. After $t_0$, they experience extended acceleration, resulting in heightened longitudinal momentum. Typically, the majority of pairs emerge during instances corresponding to local maxima of the field. Subsequently, these produced pairs undergo acceleration due to the electric field, and the gained momenta are \cite{olugh2019pair}.
\begin{equation}
q = \int_{t_{0}}^{t} e \mathbf{E}(t') \mathrm{d} t' = e \mathbf{A}\left(t_{0}\right)-e \mathbf{A}(t).
\end{equation}

The particle final momentum is determined solely by the vector potential at the time  of its creation, as the vector potential asymptotically approaches zero as  $t\longrightarrow \infty $. For instance, the peak observed in Figure \ref{2}(a)  when $ k  =  1$  at  $q\left ( t_{0}  \right )  = 0$ can be attributed to the dominant peak in the electric field at $t = 0$. As time progresses, the electric field $\mathbf{E}(t) $ decreases, leading to a reduction in the number of particles produced. However, the vector potential $ \mathbf{A}(t) $ increases simultaneously, resulting in these particles undergoing stronger acceleration. Consequently, the positions of the peaks and the patterns in the momentum spectrum are influenced by the pulse shape.

\begin{figure}[t]
\begin{center}
\includegraphics[width=\textwidth]{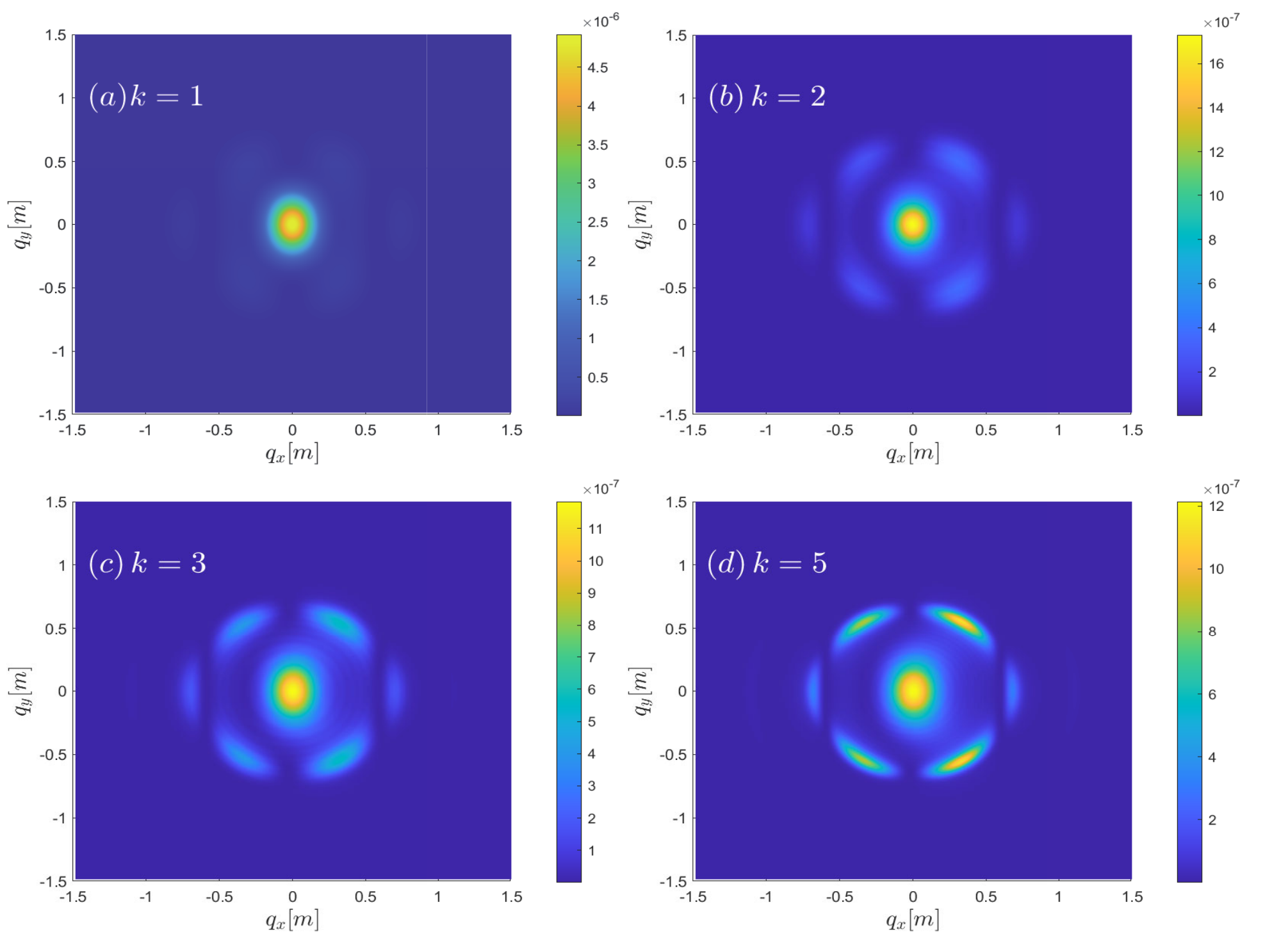}
\end{center}
\setlength{\abovecaptionskip}{-0.5cm}
\caption{Same as Fig. \ref{2} except that the falling pulse length $\tau_2 = k\tau_1$ becomes longer with $k\ge 1$. }
\label{3}
\end{figure}

Moreover, with an increased pulse length ratio $(k \geq 1)$, the momentum spectra of particles in the $(q_x, q_y)$ plane, as shown in Fig. \ref{3}, exhibit a progressive decrease in the central distribution and a gradual decrease in the peak values. Both of these phenomena are consequences of the intensifying asymmetry of the background electric field. A significant observation is the emergence of ring-like structures in the momentum spectrum. This characteristic pattern is indicative of multiphoton pair production. Since the predicted particle canonical momentum is
\begin{equation}
\left|\mathbf{q}^{*}\right|=\sqrt{\left(\frac{n \omega}{2}\right)^{2}-m_{*}^{2}},
\end{equation}
where $m_*$ and $n$ are the effective mass and absorption photon number \cite{kohlfurst2014effective}. The inner ring corresponds to the absorption of four photons, while the incomplete outer ring structure corresponds to the absorption of five photons.

\subsection{Small Frequency chirp parameter $b = 0.002$}

\begin{figure}[t]
\begin{center}
\includegraphics[width=\textwidth]{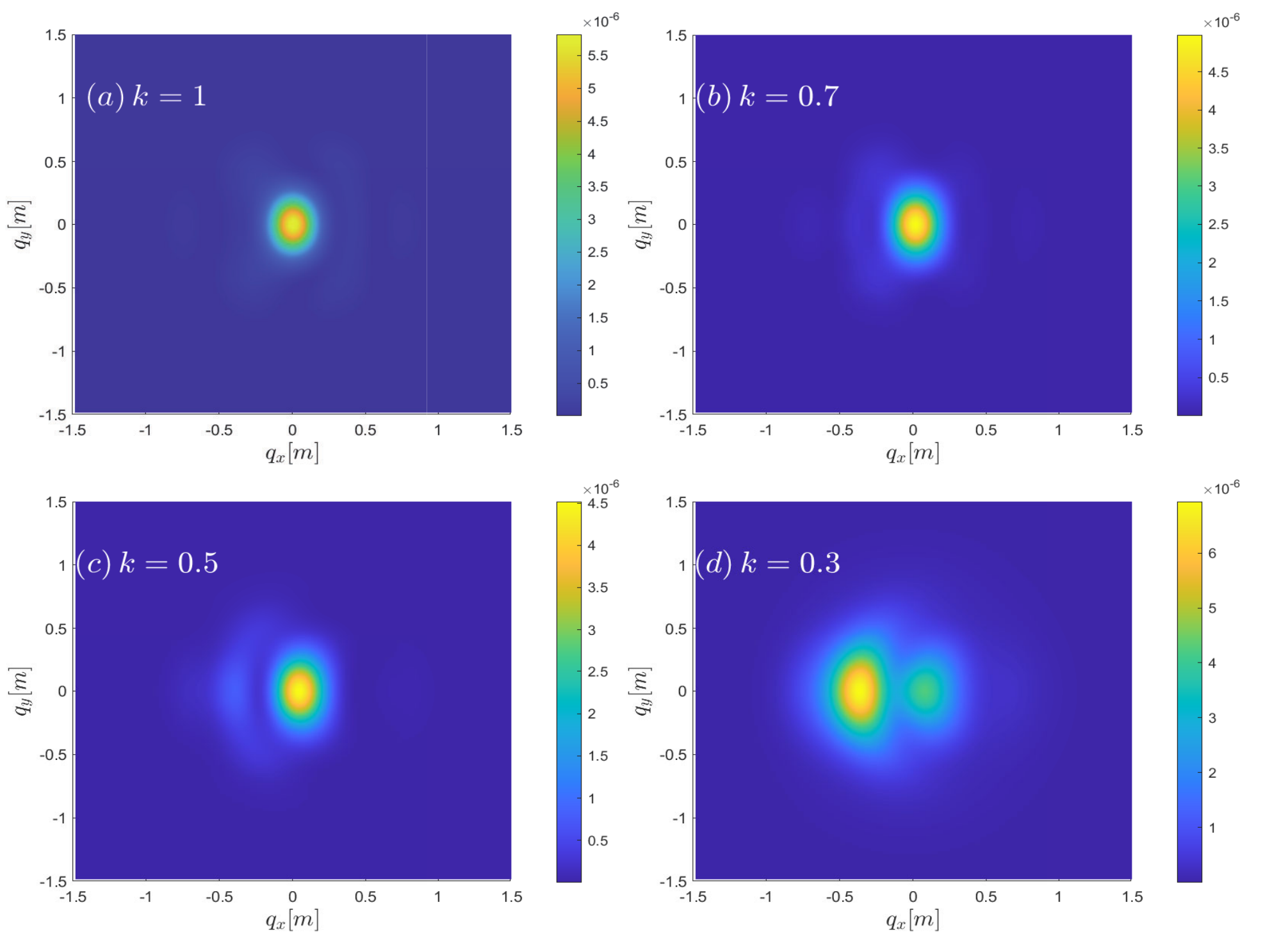}
\end{center}
\setlength{\abovecaptionskip}{-0.5cm}
\caption{Same as Fig. \ref{2} except for the small frequency chirp parameter with $b = 0.002$.}
\label{4}
\end{figure}

Considering a small frequency chirp parameter $(b = 0.002)$, Fig. \ref{4} presents momentum spectra corresponding to shorter pulse length ratios. Specifically, the peak of the momentum spectrum concentrates in the center with a value of $5.82 \times 10^{-6}$  as depicted in Fig. \ref{4}(a). This peak exhibits a slight increase compared to conditions without frequency chirp effects. Notably, the distortion of the momentum spectrum and the disruption of symmetry about the $q_x$ axis can be attributed to frequency chirp. As $k$ decreases further, two distinct momentum peaks emerge, with the left peak continuing to increase, while the right peak gradually diminishes. When $k$ is set to $0.7$, a minor increase in the momentum distribution in the negative $q_x$ region is observed, as shown in Fig. \ref{4}(b). This phenomenon is associated with pulse asymmetry. This can be attributed to the shorter falling-pulse duration $k \tau_1$,  leading to more particles being accelerated towards the negative $q_x$ direction.  At  $k = 0.5$, the momentum spectrum distinctly splits into two peaks, extending towards both positive and negative $q_x$ directions. The peak in the negative $q_x$ region is smaller, as can be seen in Fig. \ref{4}(c).  As $k$ decreases to 0.3, the peak in the negative $q_x$ region surpasses that in the positive $q_x$ region, as shown in Fig. \ref{4}(d). Furthermore, the oscillation cycles are few for shorter pulses, making the multiphoton pair production signal less clear in the spectrum due to the influence of the smaller pulse length $\tau$ on the Keldysh parameter $\gamma$.

\begin{figure}[t]
\begin{center}
\includegraphics[width=\textwidth]{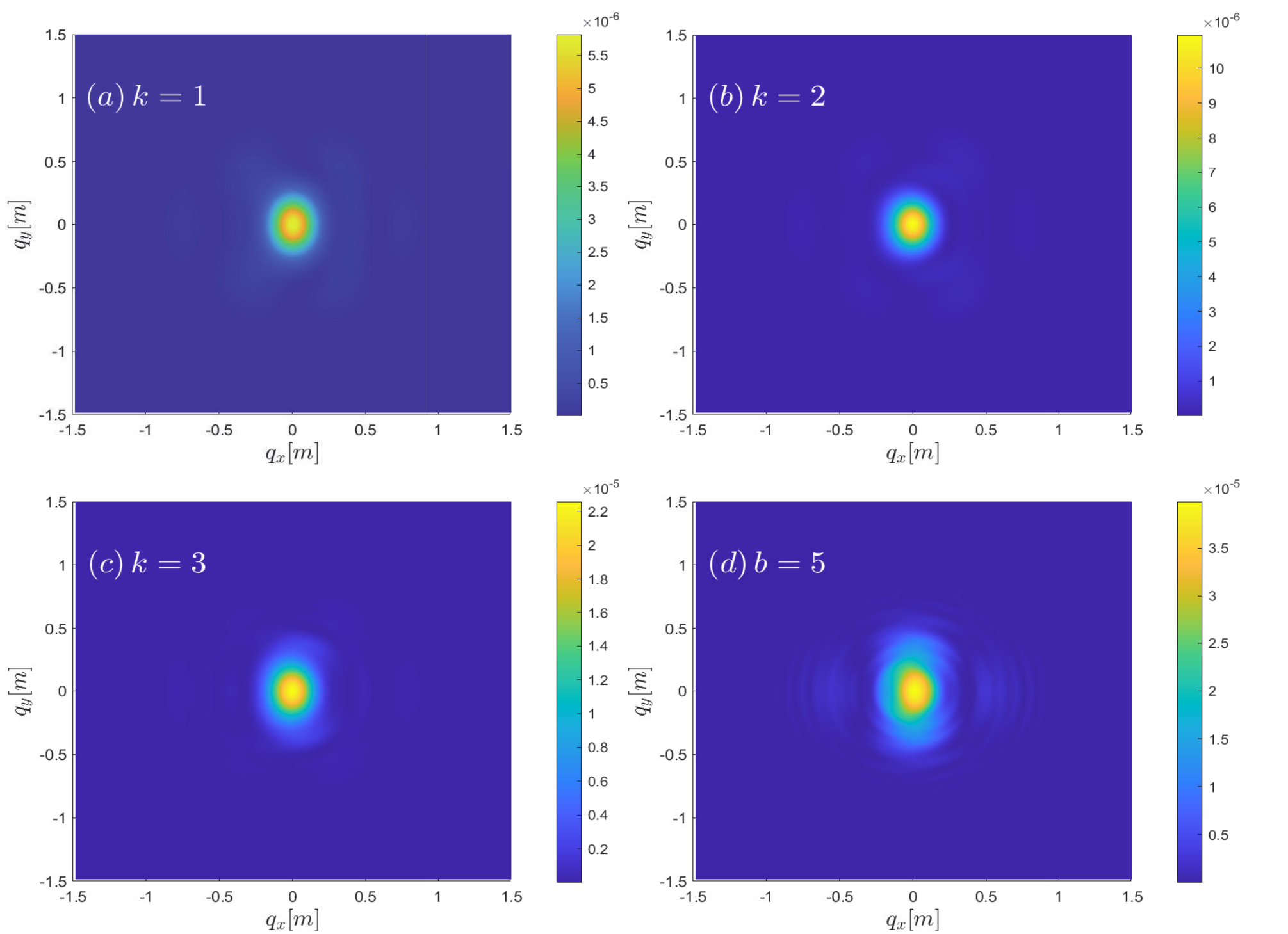}
\end{center}
\setlength{\abovecaptionskip}{-0.5cm}
\caption{Same as Fig. \ref{3} except for the small frequency chirp parameter with $b = 0.002$.}
\label{5}
\end{figure}

We consider the falling-pulse extension $(k\ge 1)$, as illustrated in Fig. \ref{5}. For an extended falling-pulse, the peak of the momentum spectrum, situated at $(0,0)$, increases with $k$. At $k = 5 $, additional incomplete ring structures emerge due to chirp effects. This can be attributed to the electric field sufficient duration and direction change during pair creation as pulse length $k \tau_1$  increases. Hence, particles might be accelerated variably, resulting in a spectrum ring structure. As pulse duration escalates with $k$, the oscillation cycles within the Gaussian envelope amplify, leading to more photons contributing to pair production.

\subsection{Large Frequency chirp parameter $b = 0.01$}

\begin{figure}[t]
\begin{center}
\includegraphics[width=\textwidth]{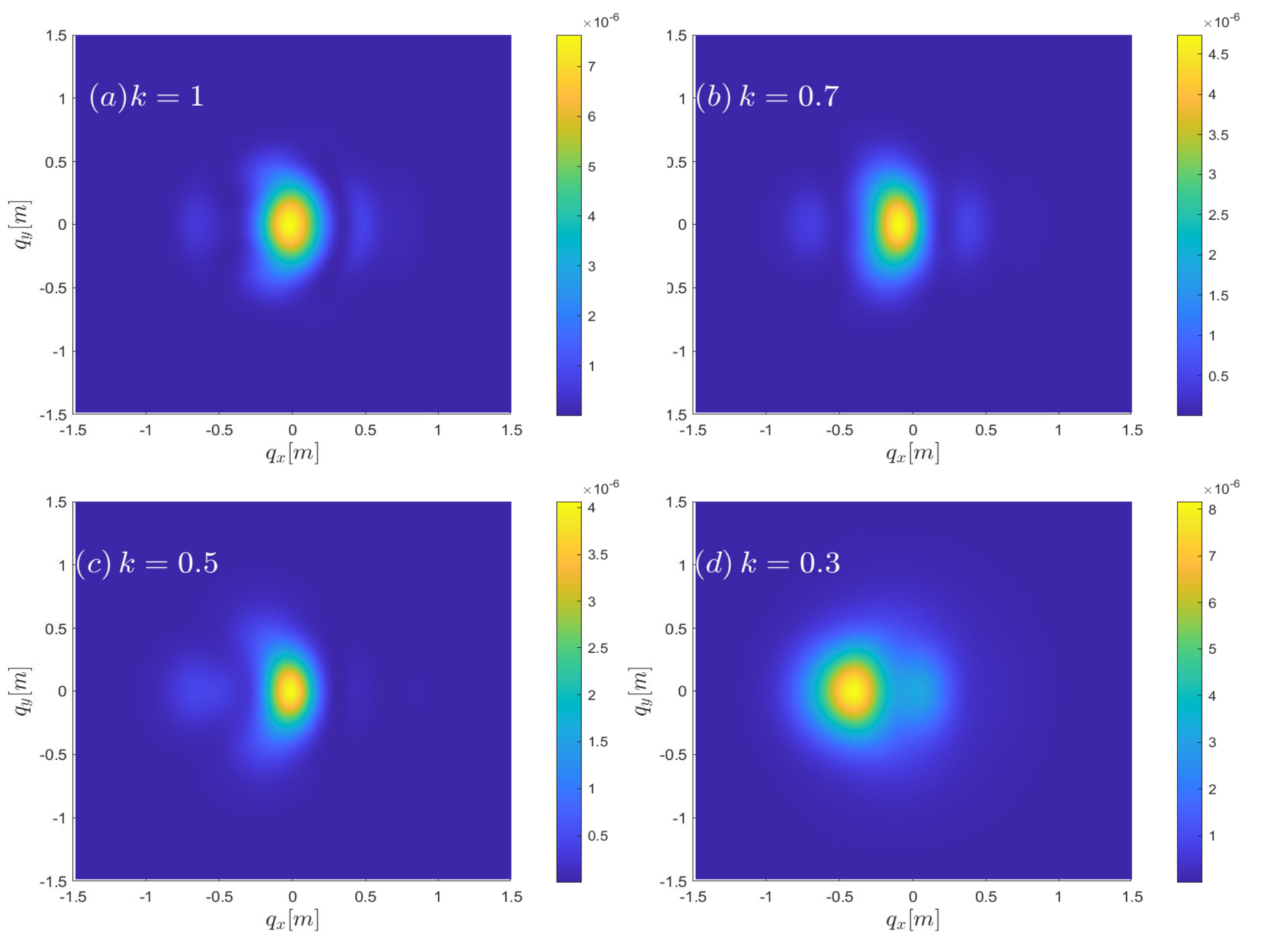}
\end{center}
\setlength{\abovecaptionskip}{-0.5cm}
\caption{Same as Fig. \ref{2} except for the large frequency chirp parameter with $b = 0.01$.}
\label{6}
\end{figure}

Considering a large frequency chirp parameter, $b = 0.01$, we delineate the momentum spectra for short pulses within the range  $0< k\le 1$ in Fig. \ref{6}. In the symmetric case $ (k = 1)$,  the momentum spectrum moves in both positive and negative $q_x$ directions, resulting in a split of the momentum spectrum, leading to the observation of three maxima, accompanied by an asymmetrical distribution, as shown in Fig. \ref{6}(a). For $k = 0.7$, the momentum spectrum becomes more focused, accompanied by a decrease in the peak value. When $k$ is set to $0.5$, the momentum spectrum takes on a jellyfish-like shape. A substantial number of particles are concentrated at the head of the jellyfish-like momentum spectrum, while the tail exhibits a lower particle distribution. At  $k = 0.3$, the jellyfish-like momentum distribution displays a different tail orientation compared to the case when  $k = 0.5$. As the pulse length ratio $k$ decreases, the number of oscillations within the envelope becomes exceedingly limited, thereby restricting the increase in effective frequency $\omega_{eff}$. Therefore, Fig. \ref{6} (d) is similar to that in  Fig. \ref{2} and \ref{4} (d).

\begin{figure}[t]
\begin{center}
\includegraphics[width=\textwidth]{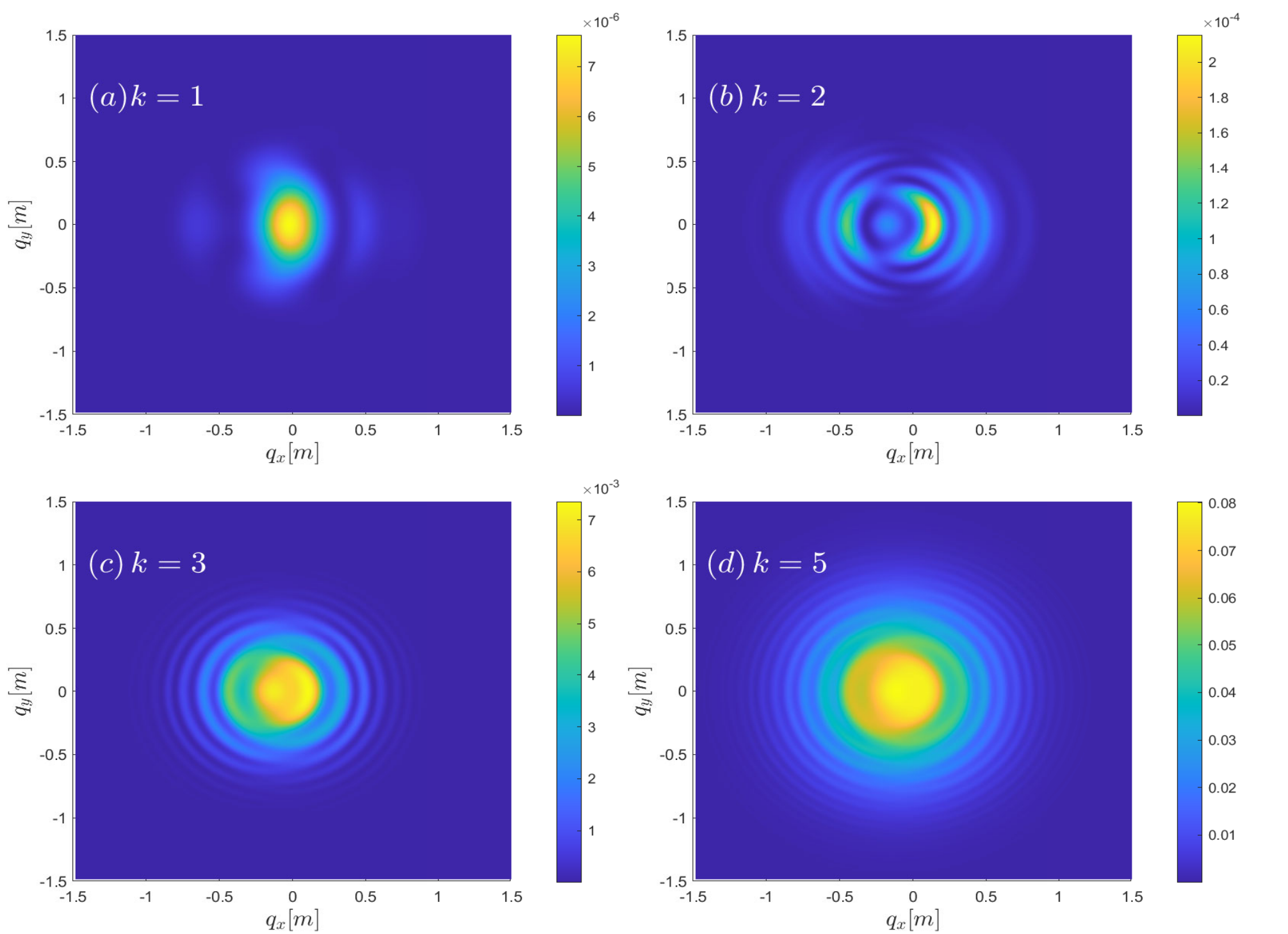}
\end{center}
\setlength{\abovecaptionskip}{-0.5cm}
\caption{Same as Fig. \ref{3} except for the large frequency chirp parameter with $b = 0.01$.}
\label{7}
\end{figure}

For a large frequency chirp parameter, $b = 0.01$, we present the momentum spectra for an extended pulse length ratio $(1\le k\le 5)$ in Fig. \ref{7}. As the falling-pulse duration $k \tau_{1}$ increases, the effect of the chirp frequency is magnified. At $k = 2$, intriguing interference patterns are observed in Fig. \ref{7} (b).  The extended duration of the falling pulse induces extensive interference, attributable to a significant number of oscillation periods within the envelope.  When $ k = 3 $, partial circular interference patterns are observed in Fig. \ref{7}(c). With increasing pulse length $k \tau_1$, the electric field persists long enough to alter its direction during the pair creation process. This results in a ring structure of the spectrum.  The asymmetry of the pulse duration contributes to the emergence of these partial circular patterns. Finally, for $k = 5$, these circular structures accumulate layer by layer, as depicted in Fig. \ref{7}(d). Particles are primarily localized at the center of the circular rings in momentum spectrum. As the ring radius expands, the momentum distribution diminishes.

\section{Number density of pair production}\label{density}

\begin{figure}[t]
\begin{center}
\includegraphics[width=\textwidth]{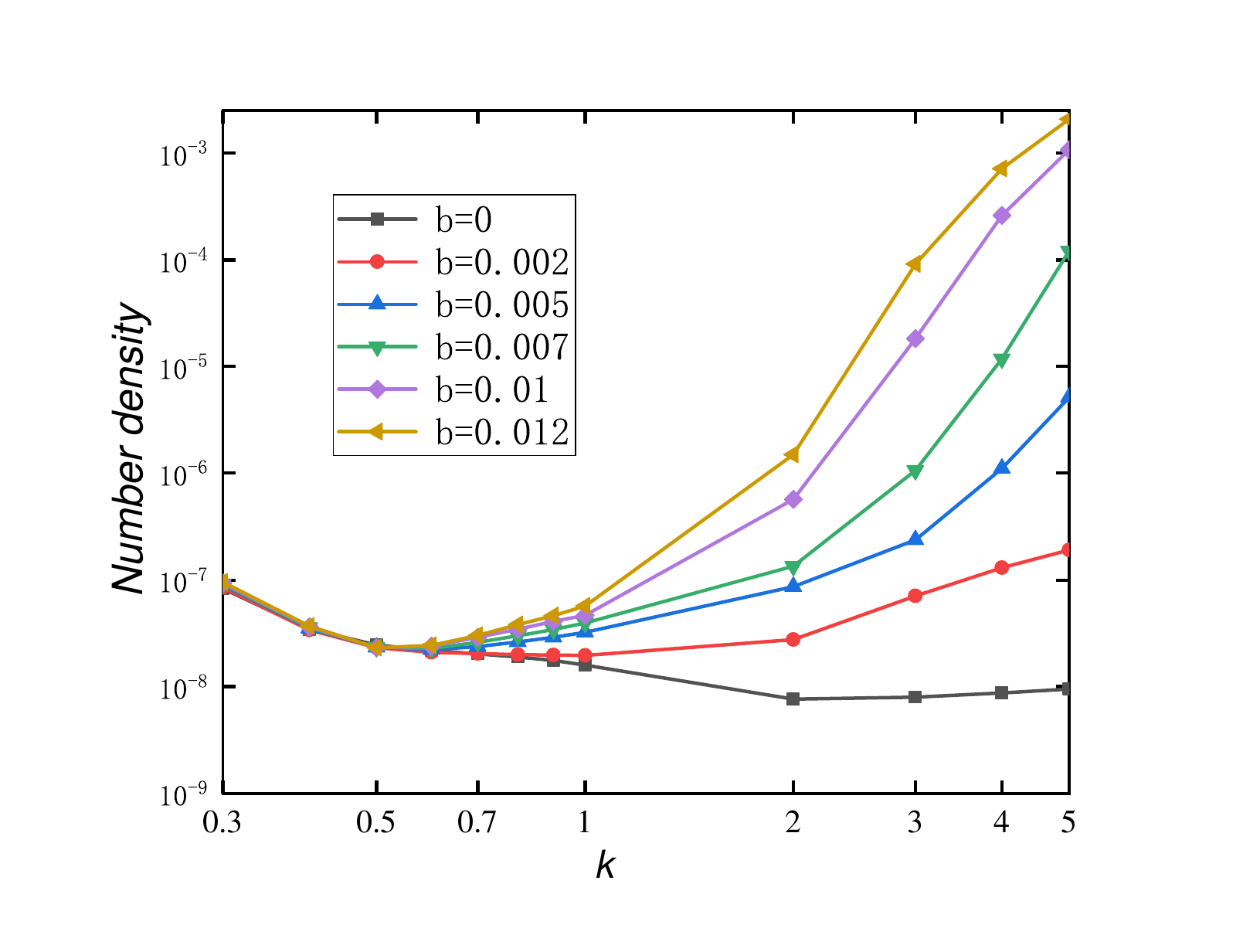}
\end{center}
\setlength{\abovecaptionskip}{-0.5cm}
\caption{Number density of particles in asymmetric electric fields with varying pulse length ratio values $k$, measured in units of $\lambda_{c} ^{-3} $. The chirp parameters $b$ considered in this research are $0, 0.002, 0.005, 0.007, 0.01$, and $0.12$. The electric field parameters chosen are $E_0 = 0.1E_{\mathrm{cr}}$, $\omega= 0.6m$, and $\tau_1$= $10/m$.}
\label{8}
\end{figure}

In this section, we present the number density of particles in chirped electric fields with the pulse ratio $k$ varying as depicted in Fig. \ref{8}.
For the chirp-free field $(b = 0)$, we observe a decrease in the particle number density as the pulse length ratio $k$ increases. Specifically, the number  density drops from $8.31 \times 10^{-8}$ for $k = 0.3$ to $9.50 \times 10^{-9}$ for $k = 5$.  As $k$ decreases, the field consists of strong pulses for $t < 0$ and weak  but with a wider frequency component (in terms of Fourier decomposition) pulses for $t > 0$ . These two semi-pulses, operating at different time scales, serve as an effective dynamically assisted mechanism, leading to an increase in the number density.
A similar phenomenon is observed in asymmetric linearly polarized pulses in Figs. \ref{7} and \ref{8} from the Ref. \cite{olugh2020asymmetric}.

For a small chirp parameter $(b = 0.002)$,  we observe an initial decline in the number density of produced particles as the pulse ratio $k$  increases. This decrease reaches a minimum at $k = 1$, corresponding to a value of $1.96 \times 10^{-8}$. When  $k> 1$, the number density begins to rise. The extension of the pulse duration is accompanied by an increase in the number of oscillation cycles, thereby enhancing the multiphoton mechanism. Consequently, number density of particles rapidly increases for larger pulse ratio $k$. Interestingly, the chirp does not necessarily enhance the number density when $b$ is small. This nonmonotonic relationship between the number density and the chirp $b$ is mainly attributed to the temporal structure of the external field.

For a larger chirp parameter, the number density of particles generated  initially decreases and then increases with the pulse length ratio $k$. With decreasing values of $k$, the dynamically assisted mechanism becomes progressively more effective, thereby enhancing the number density.  Conversely, for larger values of $k$, the pulse duration extends, resulting in an increased number of oscillation cycles within the Gaussian envelope. Simultaneously, the effective frequency $\omega_{eff}$  increases. Consequently, more photons contribute to pair production through the multiphoton absorption mechanism. For instance, in an electric field with chirp parameter $b = 0.012$, the number density reaches its minimum when $k$ is $0.5$. The two mechanisms of enhanced pair production compete, resulting in the lowest number density.

In conclusion, reducing the falling pulse length in the chirped field can potentially  increases the number density by up to  fivefold. Conversely, extending the falling pulse can  enhance the number density by one to four orders of magnitude.  Therefore, among asymmetric electric fields with varying chirp, elongating the falling pulse emerges as the most effective strategy for enhancing particle production.

\section{summary and discussion}\label{conclusion}

In this study, we investigate the effects of the asymmetric pulse shape on the momentum spectrum of created ${e^{-}e^{+}}$ pairs in strong electric fields. We considered three distinct chirp fields: chirp-free, small frequency chirp, and large frequency chirp fields. Utilizing the DHW formalism, we analyze the momentum spectrum of the generated particles. Keeping the rising pulse $\tau_1$ constant  while modifying the falling pulse $\tau_2 = k \tau_1$, leading to two asymmetric scenarios: pulse compression $(0< k< 1)$ and pulse extension ($k>1$). As the falling pulse  $k \tau_1$ compresses, interference effects gradually diminish with the decreasing pulse length ratio $k$. Shortening the pulse length induces a shift and split of the peaks. Conversely, when the falling pulse $\tau_2$ extends, an incomplete multiring structure appears for different chirped fields. The signal of multiphoton pair production becomes pronounced in the chirp-free field, while interference effects become noticeable in chirped electric fields.

We also explore the impact  of asymmetric falling pulse on the number density. The particle number density is found to be highly sensitive to both the asymmetry and chirp parameters of the pulse. When the falling  pulse length is compressed, the number density  could increase by up to fivefold.  Importantly, an elongated falling pulse leads to a significant enhancement in the number density of produced pairs by more than four orders of magnitude. The number density is influenced by  the effective dynamically assisted mechanism and the frequency chirp. For shorter pulse $(k < 1)$, we observe that the traditional standard multiphoton pair production weakens due to the Keldysh parameter $\gamma = m\omega /eE$ being modified by  another time scale $\tau$ of the pulse duration~\cite{olugh2020asymmetric}. For very small $\tau$, this implies that the oscillation number of the field encompasses  fewer cycles and subcycles, making it not strictly a complete multiphoton process. However, in this case, the number density of pairs can be increased remarkably due to the dynamically assisted mechanism. On the other hand, for the elongated pulse $k > 1$, as $k$ increases, pair production is dominated by the multiphoton mechanism with chirping.

The influence of chirp frequency on momentum spectrum and number density is relatively minor during pulse compression. The number density increases rapidly with chirp parameters in the case of pulse extension. These findings contribute to understanding the impact of key external parameters, namely pulse length and chirp parameters, and provide insight into the structure of the external pulse. While these results reveal useful information about electron-positron pair production in various chirped fields, this study is confined to multiphoton pair production. Further research is essential to investigate the effects of asymmetric pulse shapes under the Schwinger mechanism.

\textbf{Acknowledgments}

This work was supported by the National Natural Science Foundation of China (NSFC) under Grant No.\ 12375240 and No.\ 11935008. The computation was carried out at the HSCC of the Beijing Normal University.

\end{document}